\documentclass[aps,twocolumn,prl,preprintnumbers,amsmath,amssymb,superscriptaddress]{revtex4-1}

\usepackage{graphicx}
\usepackage{bm}
\usepackage{hyperref}
\usepackage{IEEEtrantools}
\usepackage{natbib}
\usepackage{float}

\restylefloat{table}

\def\cm{cm$^{-1}$}
\def\ngs{NiGa$_2$S$_4$}

\begin{document}

\title{Impact of the lattice on magnetic properties and possible spin nematicity in the S=1 triangular antiferromagnet NiGa$_2$S$_4$.
}

\author{Michael E. Valentine}
\affiliation{Institute for Quantum Matter and Department of Physics and Astronomy, Johns Hopkins University, Baltimore, MD 21218, USA}

\author{Tomoya Higo}
\affiliation{Institute for Solid State Physics, University of Tokyo, Kashiwa, Chiba 277-8581, Japan}
\affiliation{CREST, Japan Science and Technology Agency, Kawaguchi, Saitama 332-0012, Japan}

\author{Yusuke Nambu}
\affiliation{Institute for Solid State Physics, University of Tokyo, Kashiwa, Chiba 277-8581, Japan}
\affiliation{Institute for Materials Research, Tohoku University, Sendai, Miyagi 980-8577, Japan}

\author{Dipanjan Chaudhuri}
\affiliation{Institute for Quantum Matter and Department of Physics and Astronomy, Johns Hopkins University, Baltimore, MD 21218, USA}

\author{Jiajia Wen}
\affiliation{Institute for Quantum Matter and Department of Physics and Astronomy, Johns Hopkins University, Baltimore, MD 21218, USA}

\author{Collin Broholm}
\affiliation{Institute for Quantum Matter and Department of Physics and Astronomy, Johns Hopkins University, Baltimore, MD 21218, USA}
\affiliation{NIST Center for Neutron Research, National Institute of Standards and Technology, Gaithersburg, Maryland 20899, USA}
\affiliation{Department of Materials Science and Engineering, Whiting School, Johns Hopkins University, Baltimore, MD 21218, USA}

\author{Satoru Nakatsuji}
\affiliation{Institute for Solid State Physics, University of Tokyo, Kashiwa, Chiba 277-8581, Japan}
\affiliation{CREST, Japan Science and Technology Agency, Kawaguchi, Saitama 332-0012, Japan}
\affiliation{Institute for Quantum Matter and Department of Physics and Astronomy, Johns Hopkins University, Baltimore, MD 21218, USA}
\affiliation{Department of Physics, University of Tokyo, Bunkyo-ku, Tokyo 113-0033, Japan}
\affiliation{Trans-scale Quantum Science Institute, University of Tokyo, Bunkyo-ku, Tokyo 113-0033, Japan}

\author{Natalia Drichko}\email{Corresponding author: drichko@jhu.edu}
\affiliation{Institute for Quantum Matter and Department of Physics and Astronomy, Johns Hopkins University, Baltimore, MD 21218, USA}

\begin{abstract}
\ngs\ is a triangular lattice  S=1 system with strong two-dimensionality of the lattice, actively discussed as a candidate to host spin-nematic order brought about by strong quadrupole coupling.  Using Raman scattering spectroscopy we identify a  phonon of E$_g$ symmetry which can modulate magnetic exchange $J_1$ and produce quadrupole coupling.  Additionally, our Raman scattering results demonstrate a  loss of local inversion symmetry  on cooling, which we associate with sulfur vacancies. This will lead to disordered Dzyaloshinskii-Moriya interactions, which can prevent long range magnetic order. Using magnetic Raman scattering response we identify 160~K as a temperature of an upturn of magnetic correlations. The temperature below 160~K, but above 50~K where antiferromagnetic magnetic start to increase,  is a candidate for spin-nematic regime.

\end{abstract}

\date{\today}
\maketitle

\label{sec:intro}

Magnetic frustration is an active area of research that drives discoveries of new quantum phenomena~\cite{Savary2017,Starykh2015}.  While much of this work is motivated by the search for a quantum spin liquid, various novel types of spin order have been predicted and  discovered  experimentally. Exotic spin states or unsolved problems can be found even for a simplest example of a frustrated system, a triangular lattice. Nearest neighbor Hiesenberg antiferromagnetic interactions in such a system with S=1/2 result in  120$^{\circ}$ spin order. This order can be prevented by a  presence of magnetic interactions beyond nearest neighbor ones. For example, in organic molecular Mott insulators, a presence of a  ring exchange results in a spin liquid state~\cite{Holt2014}. In this work we address one example of a triangular lattice system, where the absence of long range magnetic order is not yet understood, despite much effort. This system is \ngs. It was actively discussed  in a context of nematic order,  originating from  large bi-quadratic magnetic exchange. ~\cite{Stoudenmire2009,Takano2011,Bhattacharjee2006,Lauchli2006}.

\begin{figure}
	\includegraphics[width=9cm]{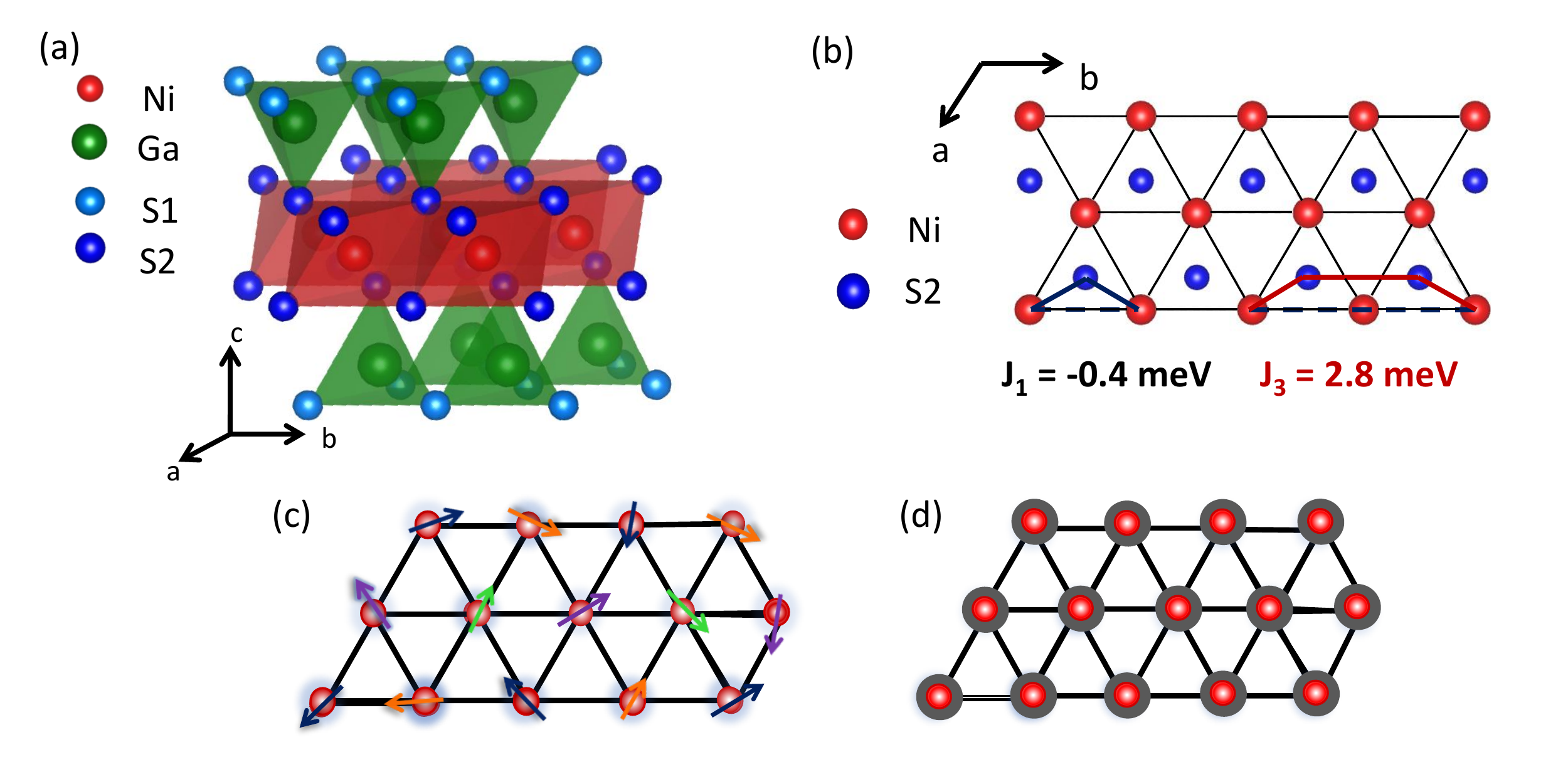}
	\caption{ (a) Crystal structure of \ngs. (b) Schematic structure of triangular lattice of Ni with J$_1$ and J$_3$ superexchanges through S2 atoms marked. (c) Schematic view of the incommensurate short range order from Ref.~\cite{Nakatsuji2005} (d) Schematic view of spin-nematic order on triangular lattice with director $d$ perpendicular to the $(ab)$ plane. Grey circles depict $S_z$=0 states. }
	\label{Fig0_basics}
\end{figure}

In \ngs\ structure, an undistorted  2D triangular lattice of Ni$^{2+}$ ($S = 1$)  is formed in the $ab$ plane from edge-sharing NiS$_6$ octahedra~\cite{Nakatsuji2005} (Fig.~\ref{Fig0_basics}a).  Additional layers of non-magnetic GaS$_4$ tetrahedra are positioned between Ni layers along $c$ axis. This  results in a highly two-dimensional structure,  with negligibly small magnetic interactions between Ni planes.

Neutron scattering \cite{Nakatsuji2005,Stock2010} results have been explained in terms of antiferromagnetic  correlations with the wave vector near (1/6,1/6,0), appearing on a triangular lattice with  a ferromagnetic  nearest-neighbor superexchange  $J_1$ = -0.4 meV, and an antiferromagnetic third nearest exchange   $J_3$ = 2.8~meV \cite{Stock2010}  (see Fig.~\ref{Fig0_basics}c). Values of exchange interactions obtained by other methods and calculations vary a lot~\cite{Pradines2018}, but agree on an importance of both  exchanges. Both competing interactions and low dimensionality of the system can lead to the  observed suppression of long range magnetic order for this system where $\Theta_W = -80$~K~\cite{Nakatsuji2005}. On lowering the temperature below 8~K  the system undergoes a gradual spin freezing~\cite{Nambu2015}.

The absence of spin order, with spin dynamics slowing down from about 50~K~\cite{Stock2010} has led to various theoretical proposals of a spin-nematic state  in \ngs \cite{Stoudenmire2009,Takano2011,Bhattacharjee2006,Lauchli2006}. Most models suggest ferronematic  order, where  full rotational symmetry of  spin is broken, and it is confined to  a certain plane defined by a ``director” that characterizes the order (Fig.~\ref{Fig0_basics} c)~\footnote{It is useful to mention, that this state is different from what is called ''nematic order'' found, for example, in FeAs.}.
Spin-nematic state is also an important starting point in the theoretical explanation for the spin freezing at 8~K, which occurs in \ngs\ in the presence of  only few per cent of sulfur vacancies~\cite{Stoudenmire2009,Nambu2009}.

Spin-nematic order is difficult to confirm experimentally.  An important step  is to identify an origin of a large bi-quadratic exchange $K_q$  necessary to establish spin-nematic order ~\cite{Stoudenmire2009,Takano2011,Bhattacharjee2006,Lauchli2006}.
As shown in Ref.~\cite{Barma1975,Tchernyshyov2011}, K$_q$ can naturally arise from a second term of an expression for magnetoelastic coupling $J=J_0$+$\frac{\delta J}{\delta r}\Delta r_{ij}$+.... and elastic response of the lattice, when a phonon with a displacement $\Delta r_{ij}$ modulates  magnetic exchange or superexchange $J$.
A bi-quadratic term K$_q$ which could arise from  a phonon modulating $J_1$ can be enhanced in \ngs\ due to  the unique near-90$^\circ$ Ni-S-Ni bond angle, as proposed  in Ref.~\cite{Stoudenmire2009}.
Here we use  Raman scattering spectroscopy and Density Functional Theory (DFT) phonons calculations to identify such a phonon.
In addition, Raman scattering  has been theoretically suggested to be a probe  of spin quadrupole excitations~\cite{Michaud2011}. To identify magnetic excitations in Raman spectra we provide a comparison to inelastic neutron scattering data.

Besides these effects directly related to the magnetism in \ngs, we present evidence for  previously undetected local inversion symmetry breaking. The effect develops independently from magneto-elastic (ME) coupling, and the amplitude of the corresponding signal increases on cooling.
We associate this lattice deformation  with sulfur vacancies~\cite{Nambu2009}   that by symmetry enable random Dzyaloshinskii-Moriya (DM) interactions.

\begin{figure}
	\includegraphics[width=9cm]{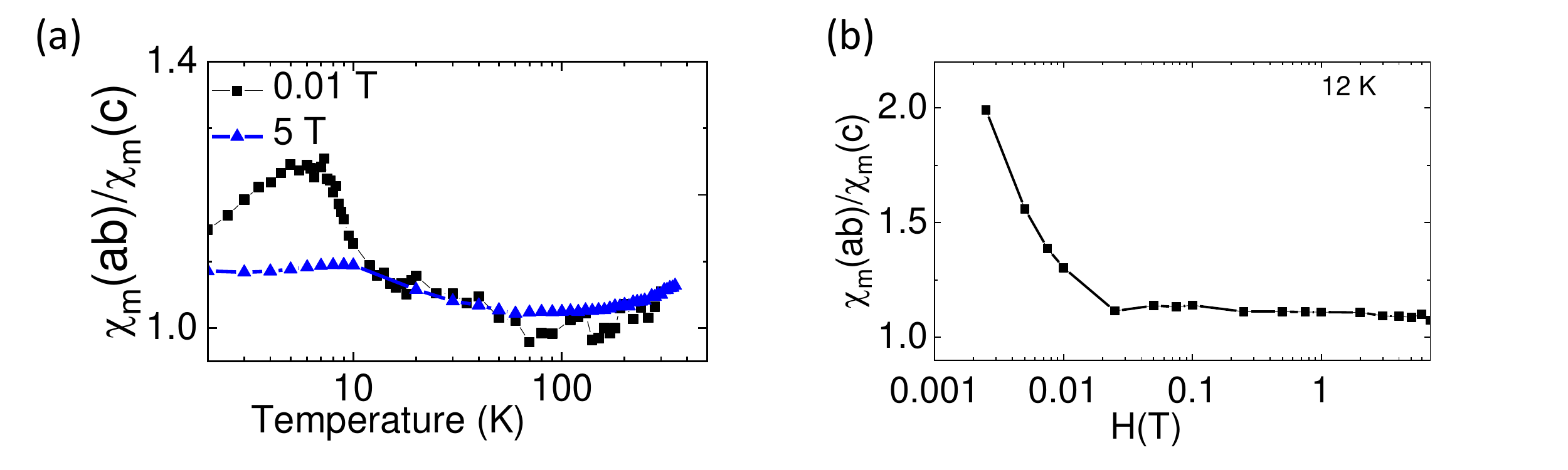}
	\caption{(a) Temperature dependence of the anisotropy of magnetic susceptibility  $\chi_m(ab)/\chi_m(c)$ at H= 0.01 T (black squares) and H=5~T (blue triangles). (b)  Field dependence of   $\chi_m(ab)/\chi_m(c)$ at 12~K.  }
	\label{fig:magnetic}
\end{figure}

The magnetic susceptibility $\chi(T)$ of \ngs\ increases on cooling, and shows flattening associated with spin freezing at temperatures below 8 K~\cite{Nakatsuji2007}.
Here we focus on the uniaxial anisotropy of the magnetic susceptibility  in magnetic field from 0.01~T to 5~T at temperatures between 2~K and 300~K, measured using a commercial SQUID magnetometer (MPMS, Quantum Design). Magnetic field H was directed parallel to the crystallographic $c$ axis ($\chi_c$) and parallel to the $(ab)$ plane ($\chi_{ab}$).  In the temperature range from 200~K to 50~ K, no considerable anisotropy $\chi_{ab}/\chi_{c}$ is observed. Below  50~K, for all measured magnetic fields  anisotropy of susceptibility   $\chi_{ab}/\chi_{c}$ increases on cooling, reaching a maximum at spin freezing temperature around 8~K (Fig.~\ref{fig:magnetic}a). This anisotropy indicates a development of in-(ab)-plane magnetic correlations. When the material enters a frozen spin state,  $\chi_{ab}/\chi_{c}$ decreases again. The maximum of the anisotropy at low temperatures is suppressed by magnetic fields above 0.025~T; a weaker magnetic anisotropy with $\chi_{ab}/\chi_{c}$(12 K) = 1.2  at 8~K is preserved up to the highest measured field of 5 T (Fig.~\ref{fig:magnetic}b).

\begin{figure}
	\includegraphics[width=9cm]{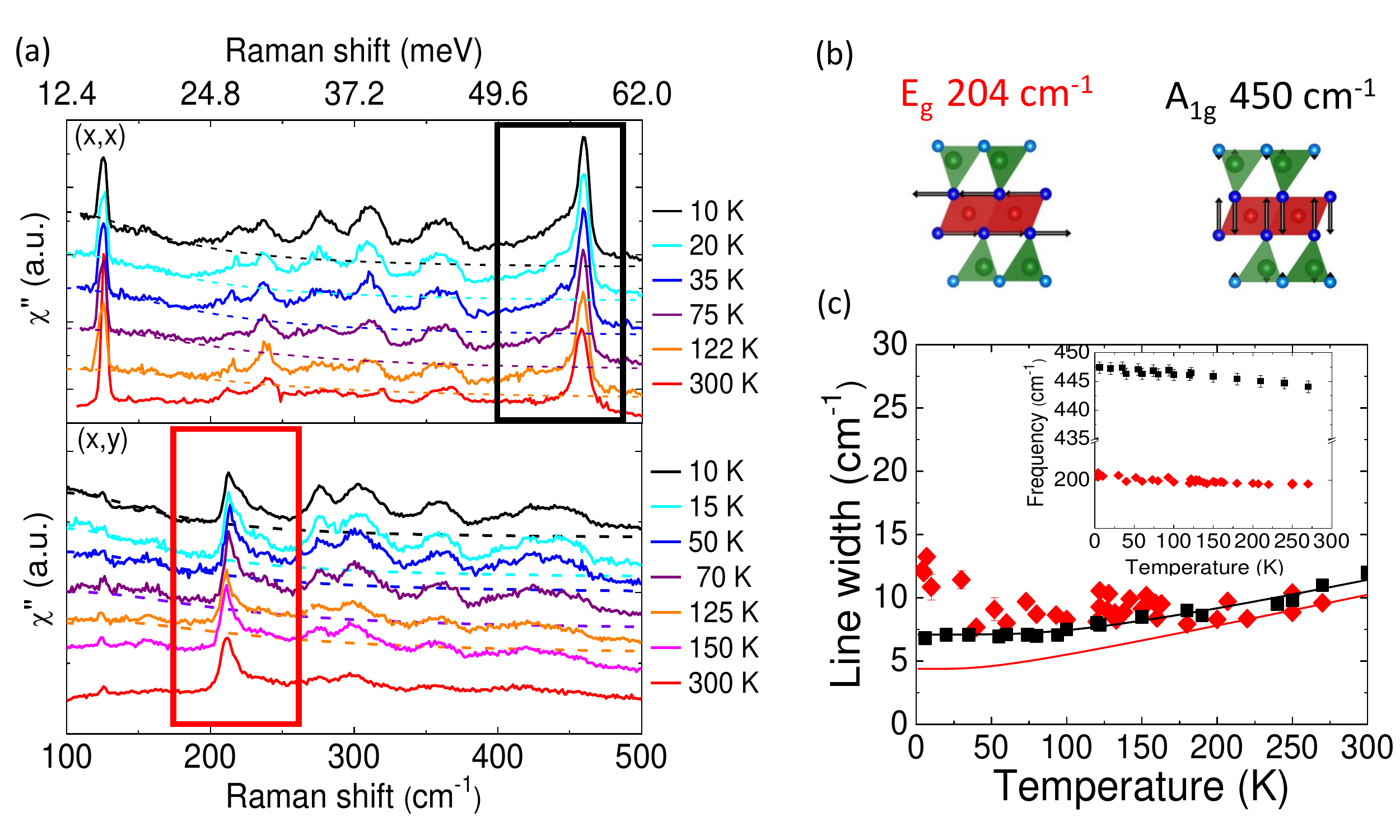}
	\caption{(a) Temperature dependence of Raman spectra of \ngs\ in $(x,x)$ and  $(x,y)$ polarizations within  the $ab$ plane. Spectra are displaced along vertical axis for clarity. Dashed lines are a guide for the eyes to show the continuum of excitations which appears in the low temperature regime. (b) Atomic displacements for E$_g$ and A$_{1g}$ phonons. (c) Temperature dependence of the line width of the E$_g$ (red) and the A$_{1g}$ (black) phonons. Note an anomalous broadening of the E$_g$ phonon below 150~K. The inset shows the temperature dependence of the phonon frequencies, which show no anomaly.}
	\label{fig:ME_coupling_A}
\end{figure}

Fig.~\ref{fig:ME_coupling_A}a shows the Raman scattering spectra of \ngs\ for temperatures between 300~K and 8 K. Raman scattering was excited by 514.5 nm (2.4~eV) line of Coherent Ar$^+$ laser, and measured using T64000 Horiba Jobin-Yvon spectrometer. Spectra were measured  from the $(ab)$ plane of the crystals, in $(x,x)$ and $(x,y)$ geometries ($x \perp y$), which correspond to the A$_{1g}$ and E$_g$ scattering channels for the $D_{3d}$ point group associated with trigonal $P\bar{3}m1$ space group of \ngs.    Raman spectra of \ngs\ comprise of a superposition of forbidden modes and magnetic excitations, in addition to Raman-active phonons. Five of six ($\Gamma_R$= 3A$_{1g}$ + 3 E$_g$) Raman-active phonons appear in the spectra as sharp intense peaks (see Supplemental Information (SI)). Here we focus on the two phonons at 206 and 450~\cm\ that involve movement of sulfur atoms S2 that form NiS$_6$ octahedra~(see Fig.~\ref{fig:ME_coupling_A}). These  are of a special interest, since S2 mediate all magnetic interactions in \ngs~ (see Fig.~\ref{Fig0_basics} b).  The sulfur S2 motion associated with these two phonons  modulates the nearest neigbor exchange interactions $J_1$ ($\delta J_1/\delta r$, where $r$ is a change of S2 coordinate) in two different ways (see Fig.~\ref{fig:ME_coupling_A}).  The A$_{1g}$ phonon at 450~\cm corresponds to the out-of-plane movement of S2 atoms, which modulates the sign of $J_1$, but preserves the symmetry of triangular lattice.
The E$_g$ phonon at 206~\cm\ on the other hand maintains the sign of $J_1$, while dynamically breaking $C_3$ symmetry of the  Ni-S2-Ni bonds.

Experimentally, we find an essential  difference in the magneto-elastic coupling for these two phonons,  demonstrated by the temperature dependence of their line widths (Fig.~\ref{fig:ME_coupling_A} c). In the absence of magneto-elastic coupling,  a phonon line width is determined by phonon-phonon scattering~\cite{Kim2012}. It follows the
general formula $\Gamma(T,\omega)=\Gamma_0+A(2n_B(\omega/2)+1)$,  where $\Gamma_0$ is a temperature independent term defined by disorder. This formula describes well the temperature dependence of the width of the $A_{1g}$ mode at 450~\cm (Fig.~\ref{fig:ME_coupling_A}c, black squares are experimental points, a solid line is the fitting curve). In contrast, the width of the $E_g$ phonon at 206~\cm\ starts to deviate from the conventional behavior at about 150~K, and increases below 50~K. Moreover, the E$_g$ phonon at 206~\cm\  shows a characteristic asymmetric  Fano line shape, which is typically a result  of an interaction between a single level of a phonon mode and an underlying continuum. The line shape is described by  $F(\omega, \omega_F, \Gamma_F, q)=\frac{1}{\Gamma_F q^2}\frac{[q+\alpha(\omega)]^2}{1+\alpha(\omega)^2}$ where $\alpha(\omega)=\frac{\omega-\omega_F}{\Gamma_F}$, and $q$ is a parameter that describes the coupling to the continuum~\cite{Zhang2015,Fano1961}.
A positive coupling parameter  $q \sim$ 4 indicates that the continuum of excitations lies above the phonon frequency (see more details in SI).

The development of spin correlations in \ngs\ on cooling is reflected in the  temperature dependence of the magnetic Raman scattering. For an ordered antiferromagnet (AF), exchange magnetic Raman scattering,  realized through the light-induced exchange of electrons between  neighboring sites of different sublattices, leads to the creation of pairs of magnons with momenta $k$ and $-k$ ~\cite{Fleury1968,Perkins08}. In a state with AF fluctuations above the ordering temperature, Raman  scattering on pairs of fluctuations with opposite momenta~\cite{Kampf1992} is  observed in the spectra as an asymmetric maximum. This allows us to follow  magnetic excitations up to high temperatures, where the Raman response of AF fluctuations will be a result of a convolution of the density of states and a line width defined by a correlation length of AF fluctuations~\cite{Caprara2005}.

\begin{figure}
	\includegraphics[width=9cm]{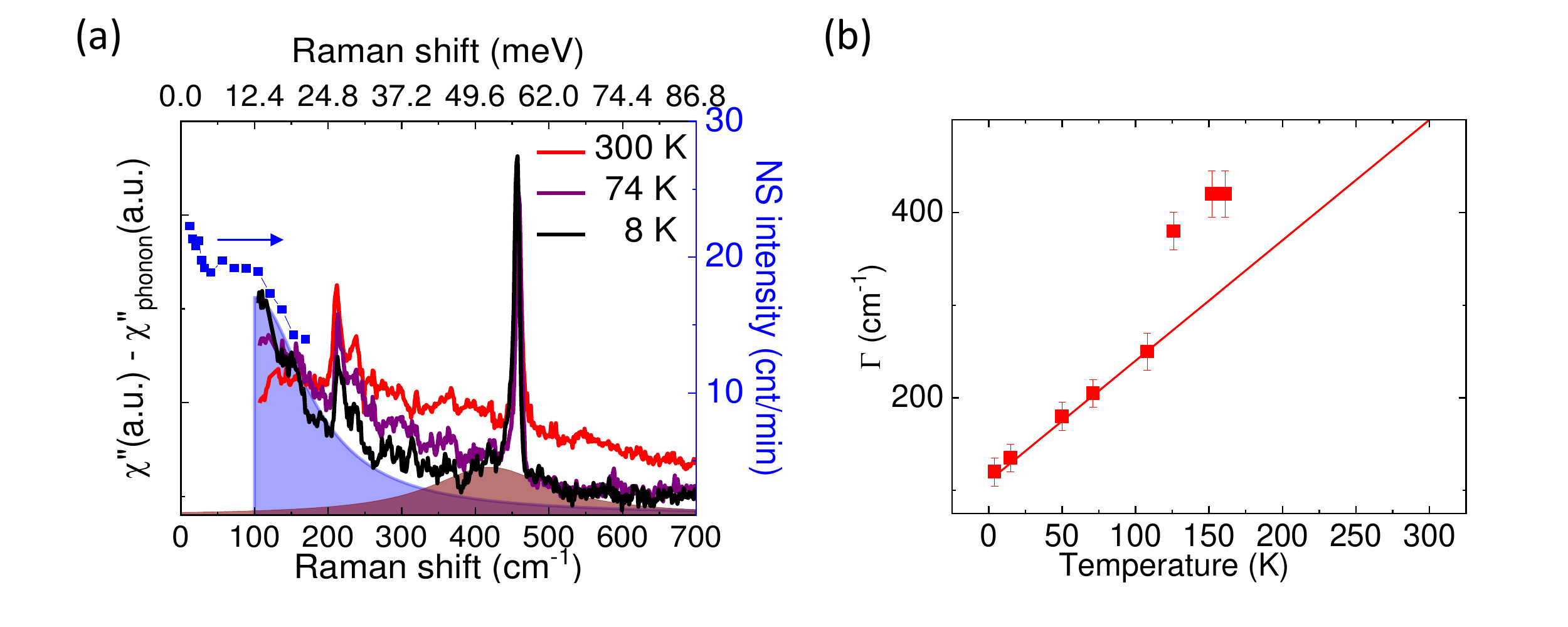}
	\caption{(a) Depolarized Raman spectra (x,x)+(x,y) at 300 K, 74 K, and 10 K with phonon that are not coupled to magnetic excitations subtracted (the S2 A$_{1g}$ and E$_g$ phonons are present in the spectra). The broad temperature dependent continuum is assigned to magnetic excitations. Blue squares show two magnon density of states calculated from  inelastic neutron scattering data taken at 2 ~K. (b) Temperature dependence of the line width of the low-frequency Raman continuum, the solid line is a linear fit. }
	\label{fig:ME_coupling_B}
\end{figure}

In Fig~\ref{fig:ME_coupling_B}a we present non-polarized Raman spectra $(x,x)$+$(x,y)$ continuum and S2  phonons, which have been isolated by subtracting  other contributions (see SI). In Fig.\ref{fig:ME_coupling_A}a   the continuum is shown by dashed lines  in the $(x,x)$ and $(x,y)$ spectra.  A low anisotropy of Raman magnetic spectra is expected for non-collinear  AF order on a triangular lattice ~\cite{Vernay2007,Perkins08}.

Fig.~\ref{fig:ME_coupling_B}a compares the low temperature Raman continuum (black line) to the low temperature (2~K) magnetic density of states $\rho(\hbar\omega)$ (blue points) inferred from neutron scattering and plotted as $\rho(E_R)$ where $E_R$ is the Raman shift. In combination these data indicate that  two-magnon excitations produce  a broad maximum around 100~\cm (12 meV). According to Raman spectra, it extends to frequencies above 500~\cm (60 meV).

Upon heating,  the low-frequency continuum starts to broaden, while the maximum does not change its position considerably. We estimate the scattering rate by a Lorenzian fit of the low-frequency maximum as indicated in Fig~\ref{fig:ME_coupling_B}a as indicated by the hatched area at 10~K. Scattering rate $\Gamma (T)$ shows linear increase with temperature up to 120~K (red curve in Fig~\ref{fig:ME_coupling_B}b). This behaviour, $\Gamma(T) = (\frac{\xi}{\xi_c})^2\Gamma(0)$, where $\xi$ is a correlation length is  expected for critical  fluctuations close to an AF transition for a simple mean field dependence of correlation length on temperature: $\xi \sim \xi_0 \sqrt{1+ T/T_c}$ ~\cite{Brenig1992, Caprara2005}.  Above 120~K  $\Gamma(T)$ starts to grow faster than linear dependence, and at temperatures above 160~K (Fig~\ref{fig:ME_coupling_B}b, red curve) the shape of the continuum changes. It becomes incoherent, with a much broader maximum at somewhat higher frequencies around 200~\cm (see SI for a more detailed temperature dependence). This change of the shape of the Raman continuum from ``critical'' fluctuations observed below approximately 160~K into a ''diffusive''~\cite{Brenig1992} background typical for any paramagnet at higher temperatures shows a change of the nature of spin-spin correlations in \ngs.

The temperature of 160~K up to which Raman scattering can trace AF fluctuations is considerably higher than 50~K, up to which NS could detect short range AF correlations in \ngs~\cite{Stock2010,Nambu2008}. Interestingly,  anisotropy of the magnetic susceptibility $\chi_{ab}/\chi_{c}$ (see Fig. ~\ref{fig:ME_coupling_B}, black points) starts to increase for T $<$ 50~K, when AF spin correlations start to develop.

Calculations for the $J_1$-$J_3$-$K_q$ model~\cite{Stoudenmire2009} suggest a lower-temperature  regime where AF spin correlations are present, while at higher temperatures  spin-nematic interactions prevail. Theoretical calculations for spin-nematic order of a S=1 system on a square lattice with bi-quadratic interactions $K_q$~\cite{Michaud2011} suggested that  Raman spectra of a quadrupole ordered state would consist of a band  at about 5$K_q$,  and  an order of magnitude weaker   excitations  bands in the region of  10$K_q$ - 12$K_q$, twice the energy of the low-frequency continuum. Spectra at temperatures between 50 and 160~K ~(Fig.~\ref{fig:ME_coupling_B}a) demonstrate a very broad continuum extending all the way to 60 meV. At 10~K we can separate a low-frequency part of the continuum with a maximum at 12 meV, and a much lower intensity continuum at frequency around 50 meV. The asymmetric shape of the E$_g$ phonon at 206~\cm, as well as slight asymmetry of   A$_g$ phonon at 450~\cm\,  suggest that they couple to the continuum states in the range of 25 to 50 meV. While these higher-frequency excitations may be the candidate for quadrupole excitations, suggested in Ref.~\cite{Michaud2011}, measurements with a higher excitation beam frequency is needed for conclusive results.

\begin{figure}
	\includegraphics[width=9cm]{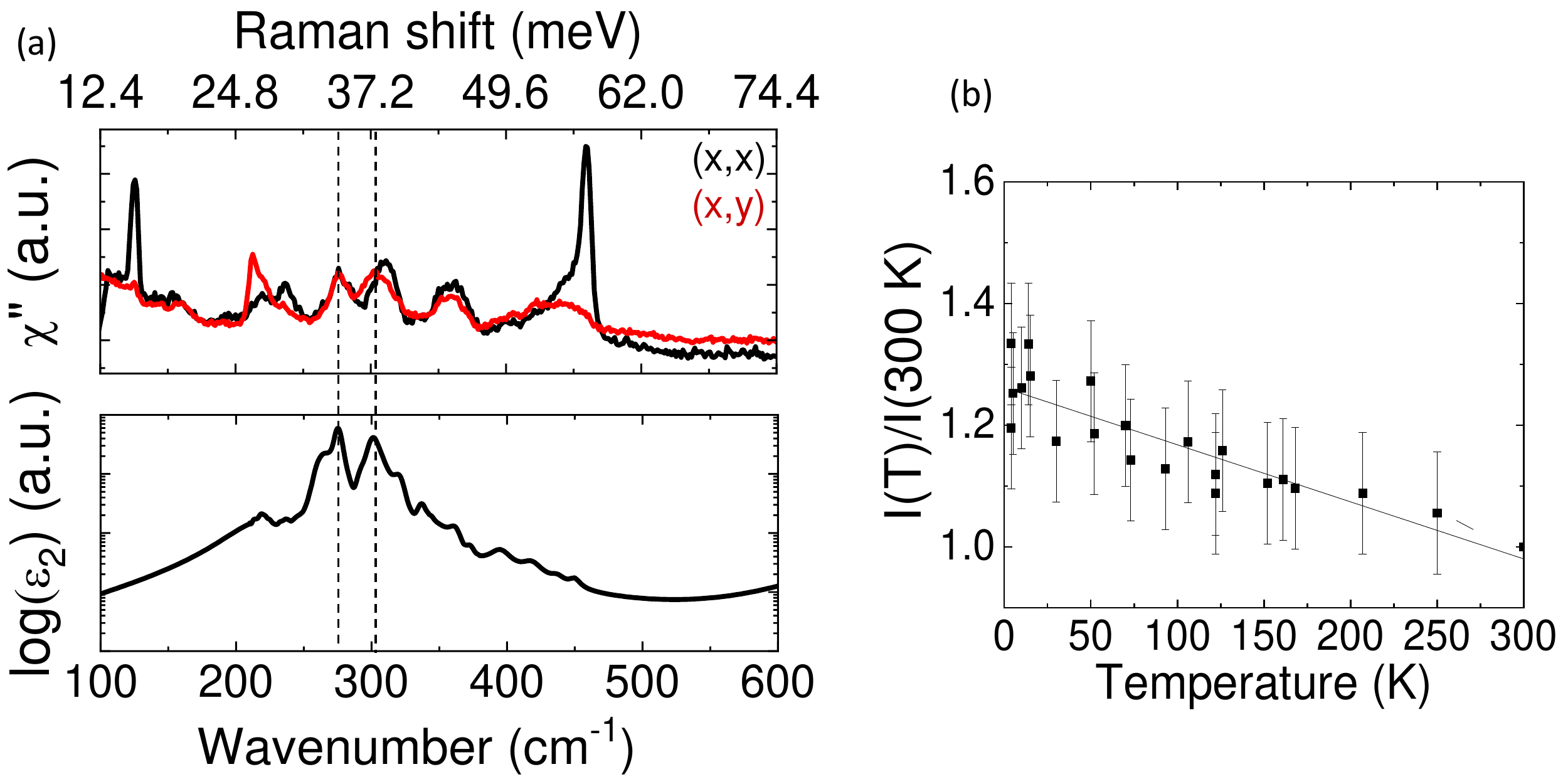}
	\caption{(a) Upper panel: Raman spectra of \ngs\ at 10~K in (x,x) (black) and (x,y) (red) channels. Lower panel: Spectra of $\epsilon_2$ for \ngs\ at 4~K. Note the presence of lines IR active phonons in both polarizations in Raman spectra. (b) Increase of the intensity of IR-active phonons in Raman spectra on cooling.}
	\label{Fig2_Inversion}
\end{figure}

Unexpectedly, and seemingly unrelated to the magnetic properties \ngs, our Raman study finds  changes of the lattice symmetry on cooling. In addition to the narrow bands of Raman-active phonons, and the continuum of magnetic excitations,  we observe   broad features which do not show distinct polarization dependence (Fig.~\ref{Fig2_Inversion}a, upper panel). Their  intensity increases linearly on cooling, as demonstrated  in Fig~\ref{Fig2_Inversion}b for the features at 270 and 296~\cm.  Comparison to $\epsilon_2$ spectra obtained from our IR reflectance measurements at T=4~K-300~K (see SI) shows, that these bands correspond to infrared-active phonons, as easily recognized for the most intense lines of  E$_u$ phonons at 270 and 296~\cm~marked on the Fig.~\ref{Fig2_Inversion}a, lower panel, with dashed lines.
$D_{3d}$ symmetry of the unit cell of \ngs\ restricts phonon modes to being either Raman or IR active~\footnote{In all inversion symmetric crystals, gerade phonons are Raman-active, and ungerade phonons are IR-active}. The appearance of  E$_u$ and A$_{2u}$ IR modes in the Raman spectrum is an indication of a loss of inversion symmetry. The absence of anomalous broadening or splitting  of  all E$_u$ and E$_g$ phonons (for details see SI) indicates that in-plane $C_3$ symmetry is unperturbed.
To the best of our knowledge, no  structural transition or crossover in \ngs\ was previously reported. However, sulfur vacancies, which are present in \ngs~\cite{Nambu2009} can break local symmetry. Local point group symmetries consistent with our observations are $C_3$ and $C_{3v}$. This means, that local deformation of the lattice around a vacancy has a dominant component parallel  to  the $c$-axis,  while preserving isotropic triangular lattice of Ni atoms in the $(ab)$ plane.

The increase of the intensity of the IR modes in the Raman spectrum upon cooling  suggests that the amplitude of these  local deformations of the lattice increases, possibly  due to anisotropic thermal contraction. At low temperatures the intensity of the IR lines becomes comparable to that of the Raman-active modes.
For local symmetry breaking  confined to unit cells with a vacancy, the estimated less than 2~\% of sulfur vacancies~\cite{Nambu2009} cannot produce such a large effect. This suggests that the symmetry breaking impact of each vacancy goes beyond a single unit cell, but does not lead to a global long range order.

To summarize, we observe two different lattice-related effects in \ngs. One is magneto-elastic coupling, the other is local loss of inversion symmetry. Both can produce considerable effects on the magnetic properties of \ngs.

The $E_g$ phonon at 206~\cm\ shows evidence of magneto-elastic coupling. It can modulate $J_1$, and  lead to a bi-quadratic exchange $K_q$. DFT calculations calculation are needed to estimate the size of the effect.

The local loss of inversion symmetry due to vacancies can be an additional factor preventing magnetic ordering.  According to  Ref.~\cite{Nambu2009},  S1 sites, which are not directly involved in superexchange,  are more suspectable to sulfur loss. The distortion of the lattice around a vacancy can  randomly change chemical bonds that determine $J_1$ and $J_3$. Additionally, the loss of inversion symmetry allows for Dzyaloshinskii–Moriya interactions, which were not previously discussed  for \ngs, but which can also affect magnetic ordering. Besides, we observed a reduction of in-plane anisotropy $\chi_{ab}/\chi_{c}$ below the spin freezing temperature. Deformation of the lattice along the $c$-axis could lie behind this  effect.

In agreement with heat capacity, Raman scattering distinguishes three distinct temperature regimes for \ngs:

In the temperature range between 300 and 160~K, no magneto-elastic coupling or antiferromagnetic fluctuations related to low-temperature magnetic state are observed in the Raman scattering.  Between approximately 160 and  50~K, the magnetic susceptibility is isotropic, but magnetic Raman scattering identifies the development of local spin correlations.  Magneto-elastic coupling is observed for E$_g$ phonon. This points on  a spin-nematic regime, but with director that is not pinned to the $c$-axis.

Below 50~K, short range dynamic spin correlations are identified by neutron scattering ~\cite{Stock2010}, with correlation length that increases on cooling, in agreement with further narrowing of the Raman low-frequency magnetic continuum. In this regime the easy-plane anisotropy of the magnetic susceptibility increases as well. Interestingly, magneto-elastic coupling for the E$_g$ phonon which modulates  $J_1$ also increases in this regime.  Our measurements of the field dependence of magnetic anisotropy show that application of only 0.025~T is enough to suppress the easy plane anisotropy, while according to  NS experiments~\cite{Stock2010} a much higher field of 5~T is needed to suppress interplane AF correlations.

The authors are   thankful to L. Balents, F. Vernay, O. Tchernyshev, and  H. Chen useful discussions, and to N. P. Armitage for a possibility of IR measurements in his lab and useful discussions. This work was supported as part of the Institute for Quantum Matter, an Energy Frontier Research Center funded by the U.S. Department of Energy, Office of Science, Basic Energy Sciences under Award No. DE-SC0019331. Work in Japan was  supported by Grants-in-Aids for Scientific Research on Innovative Areas (15H05882 and 15H05883) from the Ministry of Education, Culture, Sports, Science, and Technology of Japan, by CREST(JPMJCR18T3), Japan Science and Technology Agency, and by Grants-in-Aid for Scientific Research (16H02209, 17H06137)  from the Japanese Society for the Promotion of Science (JSPS). ND thanks ISSP for hospitality.
\\

\bibliography{./ngs2020}

\section{Supplemental Material}

\label{sec:exp}

\subsection{ Crystal growth and preparation for measurements}

 \ngs\ single crystals were grown using the method outlined in Ref.~\onlinecite{Nambu2008synthesis}. The resulting crystals are thin plates with the most developed surface parallel to $ab$ plane measuring up to 3 mm by 3 mm and thickness 10~$\mu$m.

In preparation for measurements, the samples were cleaved to ensure optimum concentration of sulfur in the probed samle. Raman scattering showed reproducible data for spectra measured from cleaved surfaces of different crystals. Magnetic susceptibility measurements showed the transition into the spin frozen state at 8~K, which corresponds to the best quality crystals with less than 4~\% sulfur vacancies~\cite{Nambu2009}.

\subsection{Neutron scattering measurements}
The neutron scattering experiment was carried out on the MACS instrument at the NIST Center for Neutron Research. 19 single crystals of $\rm NiGa_2S_4$ with total mass $\sim 1$~g were co-aligned in the $(HK0)$ plane for the measurement. A final neutron energy of $E_f=3.6$~meV was used for measurements with energy transfer $E\leq 4.5$~meV, while $E_f=5$~meV was employed for measurement with $E\geq4.5$~meV. All measurements were normalized to the monitor counts that corresponds to 1 minute counting with incident neutron energy $E_i=3.6$~meV. An aluminum sample with similar mass to the sample holder as well as the empty cryostat were measured separately at all experimental settings for background subtraction. The one-magnon density of states $\rho_1(E)$ is calculated by integrating the neutron scattering intensities measured at $T=2$~K over a common region in the second Brillouin zone, which is accessible  for all energy transfer. Under the assumption that the crystal momentum remains a good quantum number, the two-magnon density of states probed by Raman scattering $\rho_2(E)$ is related to the one-magnon density of states by $\rho_2(E) \sim \rho_1(E/2)$, which is found to agree well with the Raman measurements.

\subsection{Raman scattering spectroscopy}

The crystals of \ngs\  are shaped as thin plates, thus  spectra in $(c,c)$ polarization ($(z,z)$ scattering channel) were measured only at room temperature using a micro-Raman setup, while spectra in the $(a,b)$ plane were measured down to 4~K using a combination of macro and micro-Raman setups. Spectra in the $(a,b)$ plane were measured from freshly cleaved surfaces.

The main Raman scattering data were obtained using a macro-Raman setup based on a Jobin-Yvon T64000 Raman spectrometer in a single monochromator configuration in a pseudo-Brewster's angle scattering geometry. Measurements  at temperatures  from 300~K to 4~K in the frequency range between 100~\cm\ and 600~\cm\  were performed using a 514.5~nm line of an Ar$^+$-Kr$^+$ Coherent laser for excitation. Laser power did not exceeding 4~mW for a laser probe of $\sim 50 \times 100~\mu$m to avoid overheating of the sample. The sample was attached to the cold finger of Janis ST500 cryostat. Additional room temperature measurements  of \ngs\ crystals in $(c,c)$ polarization in the 100~\cm\ to 600~\cm\ spectral range were obtained using  the Jobin-Yvon T64000 Raman spectrometer  equipped with Olympus microscope with the spot diameter of 2~$\mu$m.  Phonon spectra in the frequency range from 10~\cm\ to 600~\cm\ with resolution 5~\cm\ were obtained using a Jobin-Yvon U1000 spectrometer equipped with a photomultiplier detector, in a pseudo-Brewster's angle scattering geometry.   514.5~nm line of an Ar$^+$ laser was used as excitation light with a beam size on the sample of $\sim 50~\mu$m $\times 100~\mu$m. Low temperature measurements were performed using a custom built Janis cold finger cryostat.

All presented spectra are normalized by a first-order Bose-Einstein thermal factor $n(\omega) + 1$.

The \ngs\ structure is described by the trigonal $P\bar{3}m1$ space group, which corresponds to the  $D_{3d}$ point group symmetry with the following Raman tensors:

\begin{align}
	A_{1g} &=
	\begin{pmatrix}
		a & 0 & 0 \\
		0 & a & 0 \\
		0 & 0 & b
	\end{pmatrix} \nonumber\\
	E_g &=
	\begin{pmatrix}
		c & 0 & 0 \\
		0 & c & d \\
		0 & d & 0
	\end{pmatrix},
	\begin{pmatrix}
		0 & -c & -d \\
		-c & 0 & 0 \\
		-d & 0 & 0
	\end{pmatrix}. \nonumber
	\label{eqn:tensors}
\end{align}

Based on these Raman tensors, the intensities for the different measured polarizations of $e_i$ electrical vector of the excitation light and $e_s$   of the Raman scattered light can be decomposed as

\begin{align}
	I_{xx} = |a|^2 + |c|^2	\nonumber, 	I_{xy} = |c|^2,		\nonumber 	
	I_{zz} = |b|^2		\nonumber
\end{align}

where $x \perp y$ denote $e_i$ and $e_s$ lying in the $(ab)$ plane, and $z$-polarized light is parallel to the out of plane $c$ axis. Discrepancies between the theoretical   and the observed polarization dependence  can be attributed to the depolarization that occurs at the crystal surface in all non-backscattering geometries.

{\bf Raman data and data analysis:}

{\bf DFT calculations} The frequencies of the phonons at the $\Gamma$ point of the BZ and the respective atomic displacements  were calculated using density functional theory (DFT) calculations  with Quantum Espresso software with the PHonon package~\cite{Giannozzi2009} based on the average structure determined by x-ray diffraction measurements~\cite{Nambu2009}. A generalized gradient approximation was used for the exchange-correlation in the energy functional.

\begin{figure}
  \centering
  \includegraphics[width=9cm]{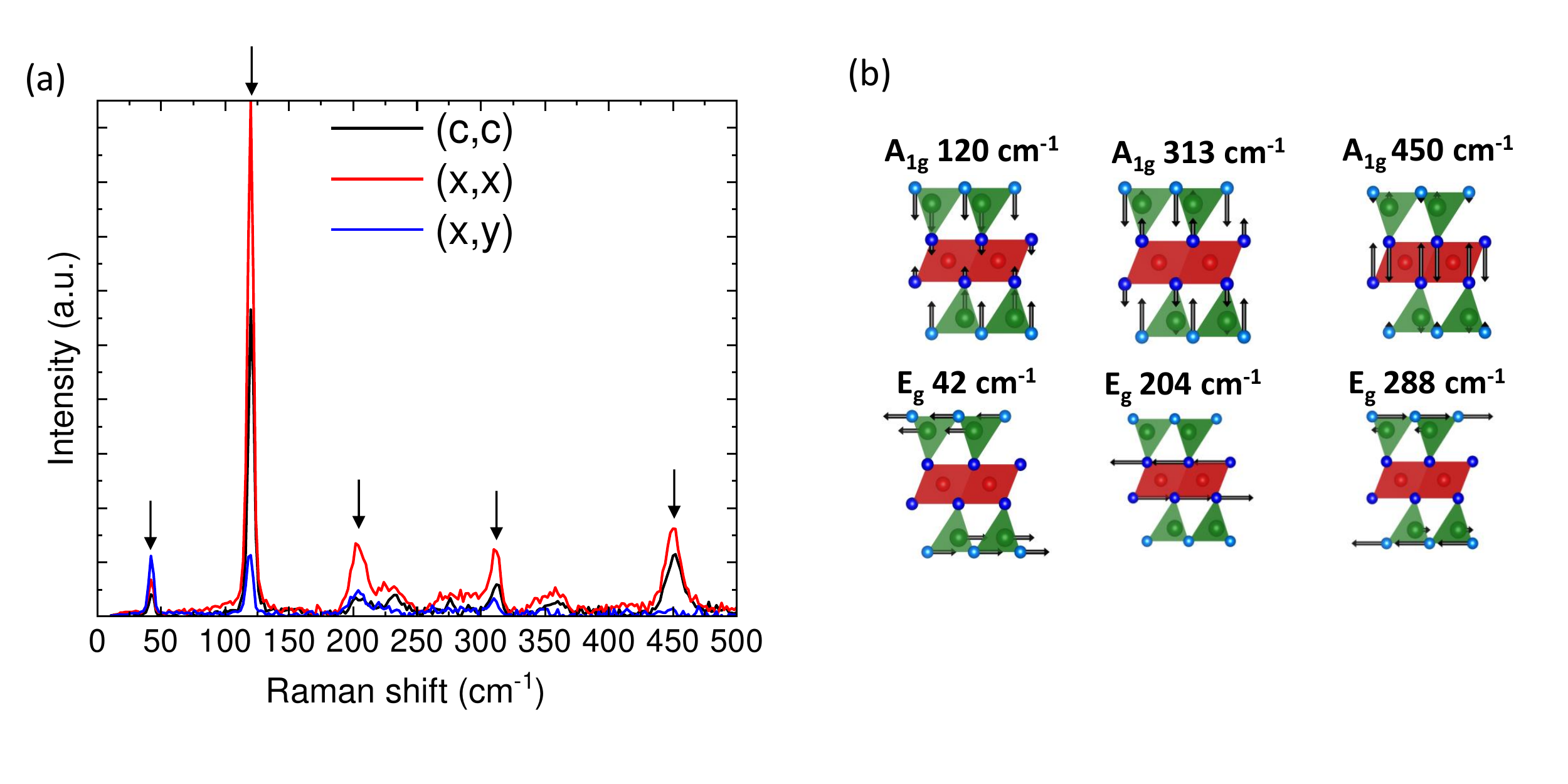}\\
\caption{Raman phonon spectra of \ngs\ at room temperature measured in 10-600~\cm\ spectral range. $(x,x)$ and $(x,y)$ are polarizations randomly oriented within the $(ab)$ plane.  $(z,z)$ polarization measured using micro-Raman setup. Five Raman-active phonons are marked with arrows. Broad low-intensity features correspond to Raman-inactive vibrations, activated due to a local loss of inversion center.  (b) Displacement of atoms for the Raman-active vibrations of \ngs\ calculated by DFT. Red are Ni-centered edge-sharing NiS$_6$ tetrahedra with S2 atoms shown in blue. Green are Ga-centered tetrahedra with S1 atoms shown in light blue.}\label{fig:rt}
\end{figure}

{\bf Assignment of pohonons in experimental spectra} The observed phonon frequency assignment is based on  our DFT calculations.
Of six Raman-active (A$_{1g}$ and E$_g$) \ngs\ phonons (Table~\ref{tab:reps}) we observe five (Fig.~\ref{fig:rt}). Another $E_g$ mode expected at 285~\cm\  is observed as a very weak band in $(x,y)$ spectra.   Likely it is more intense for $\chi''_{xz}$, $\chi''_{yz}$  ($d$ in the above Raman tensors), for which no measurements were performed. Frequencies of all the phonons observed in IR and Raman spectra and their assignment based on our DFT calculations are listed in  Table~\ref{tab:phocomp}. Calculated frequencies show a good agreement with the experiments exept for the highest frequency A$_{1g}$ phonon, which is observed at frequencies significantly  higher than the calculation suggests.

\begin{table}
	\caption{Wyckoff positions and $\Gamma$-point representations for \ngs, including  $A_{2u} + E_u$ acoustic modes. $A_{1g}$ and  $E_g$ phonons are Raman active}
	\label{tab:reps}
	\begin{tabular*}{0.45\textwidth}{@{\extracolsep{\fill}} c c c}
		\\
		\hline\hline
		Element	& Wyckoff position	& $\Gamma$ representation	\\ \hline
		Ni	& 1b			& $A_{2u} + E_u$		\\
		Ga	& 2d			& $A_{1g} + E_g + A_{2u} + E_u$	\\
		S1	& 2d			& $A_{1g} + E_g + A_{2u} + E_u$	\\
		S2	& 2d			& $A_{1g} + E_g + A_{2u} + E_u$	\\ \hline\hline
	\end{tabular*}
\end{table}


Phonon Raman spectra in the frequency range between 10 and 200~\cm\ are presented in Fig.~\ref{fig:R}a. They show two lowest lying phonons, E$_g$ at about 42~\cm, and A$_{1g}$ at about 120~\cm. Note that E$_g$ phonon is not broadened on cooling, in contrast to the $E_g$ at 206~\cm.

\begin{figure}
  \centering
  \includegraphics[width=9cm]{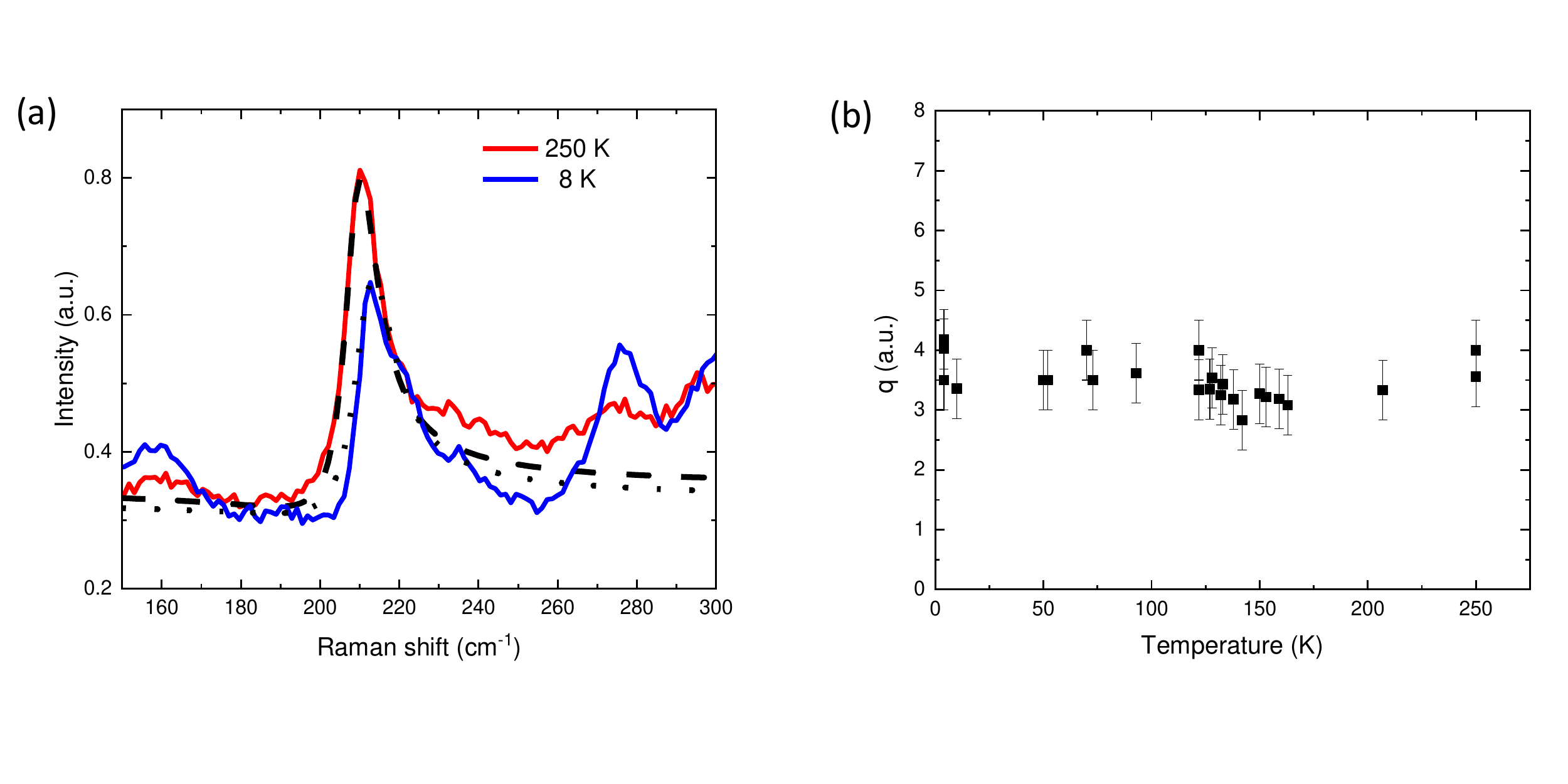}\\
\caption{(a) Fano line shape of the E$_g$ phonon (206~\cm) at 250~ K and 8 ~K. Black dotted lines show fitting curves.    (b) Temperature dependence of Fano coupling parameter $q$.}\label{fig:q}
\end{figure}

{\bf Magneto-elastic coupling} Coupling of the  E$_g$ phonon at 206~\cm\ to a continuum of excitations is described by the Fano formula  $F(\omega, \omega_F, \Gamma_F, q)=\frac{1}{\Gamma_F q^2}\frac{[q+\alpha(\omega)]^2}{1+\alpha(\omega)^2}$~\cite{Zhang2015}. We show in the main part of the text, that  upon cooling, the  width $\Gamma_F$ of the phonon obtained by least squares fitting of the spectrum increases. The temperature dependence of the width follows the growth of intensity of the continuum of magnetic excitations. In this formula, $q$ is an empirical coupling parameter, the value of which describes the strength of coupling, that depends on the spectral position of the continuum relative to the phonon~\cite{Fano1961}. The positive value of $q$ suggests that the continuum with which the E$_g$ phonon is interacting lies at frequencies above that of the phonon. The parameter  $q$  has a very weak temperature dependence (Fig.~\ref{fig:q} b), which indicates that continuum stays at constant frequencies in the whole measured temperature range.

\begin{figure}[h]
  \centering
  \includegraphics[width=9cm]{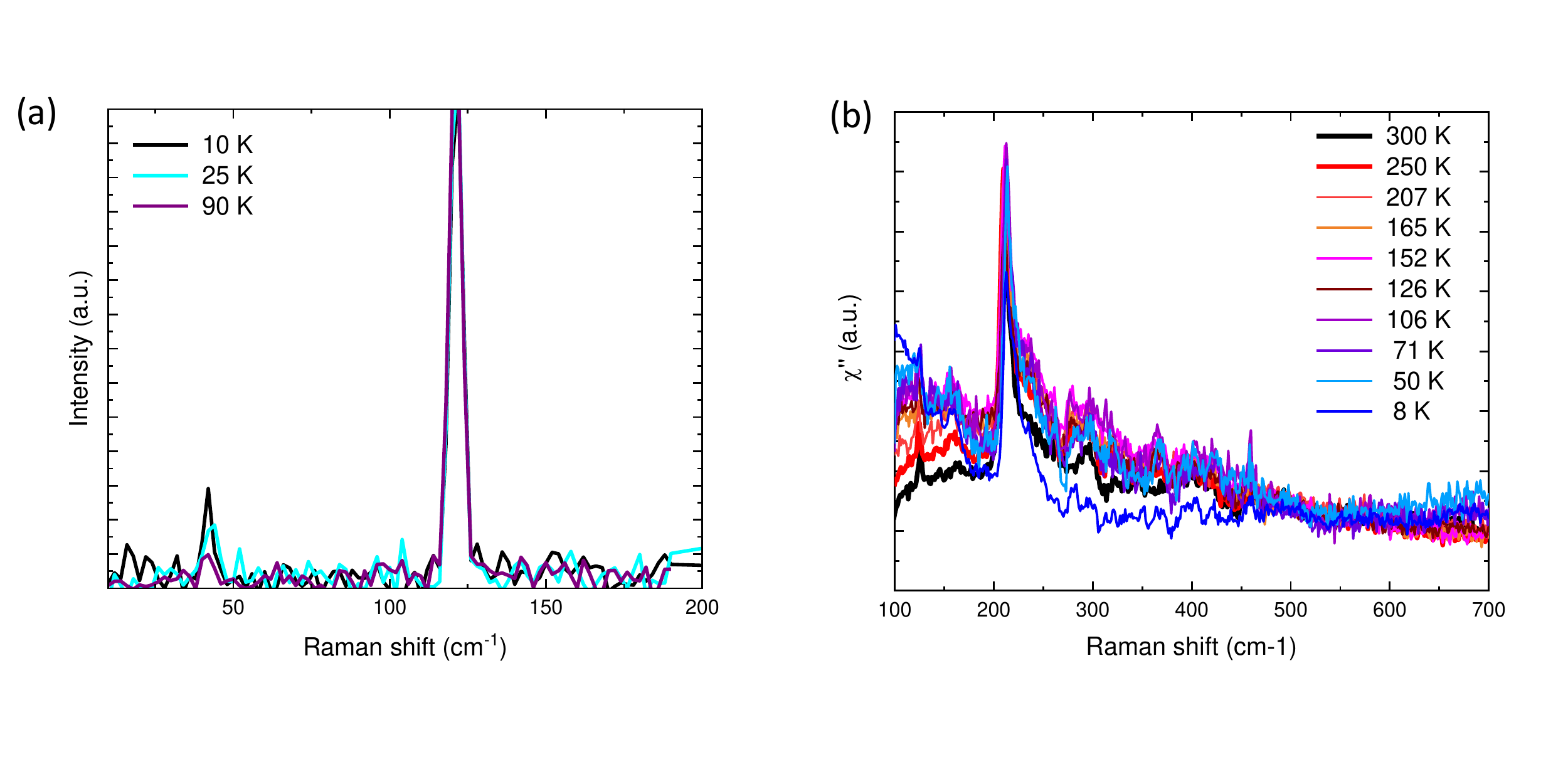}\\
\caption{(a) Raman phonon spectra of \ngs\ in the spectral range between 10~\cm\ and 200~\cm\ taken at 10~K, 25~K, and 90~K. Note the absence of broadening for the E$_g$ phonon at 42~\cm. (b) Temperature dependence magnetic Raman scattering  of \ngs\ in $(x,y)$ scattering channel}\label{fig:R}
\end{figure}

{\bf Raman-forbidden phonons subtraction}

To analyse the continuum of magnetic excitations detected in  Raman scattering spectra, we subtracted a contribution of the spectra related to Raman-forbidden phonons ($\chi''_{phonon} (\omega)$), which appear in the spectra due to local symmetry breaking, from the total experimentally measured Raman scattering $\chi''(\omega)$. As clearly observed in Fig.~\ref{fig:ME_coupling_B}a, upper panel, this contribution is isotropic, in contrast to the narrow bands of Raman-active phonons. This isotropic contribution $\chi''_{phonons} (\omega)$ was fitted with a sum or Lorenzian spectral shape, and subtracted from the spectra at each temperature: $\chi''(\omega)_{magnetic}$ = $\chi''(\omega)$-$ k(T)\chi''(\omega)_{phonon}$.

In addition to the data in Fig.~\ref{fig:ME_coupling_B}a, in Fig.~\ref{fig:R}a we present a temperature dependence of the magnetic Raman continuum measured in $(x,y)$ configuration, obtained by this subtraction.

\subsection{Low temperature IR spectra}

{\bf IR spectroscopy} A Bruker Fourier transform infrared (FTIR) spectrometer with a bolometer detector was used to obtain  the reflection infrared spectrum across an energy range from  150 \cm\ to 600 \cm\ with a resolution of 2~\cm. Spectra were measured  with polarization of light $E \parallel (ab)$, where   only modes with $E_u$ symmetry are observed. The absolute reflectance values were obtained by referencing the sample spectra to that of a sample with a gold film evaporated on its surface.  Absorbance spectra were obtained from reflectance using  Kramers-Kronig transformation.  For measurements from 4~K to 300~K a cold finger cryostat  Janis ST300 was used.  The  results are in general agreement with  infrared measurements on powder \ngs\ samples~\cite{Lutz1986}.

\begin{figure}[h]
  \centering
  \includegraphics[width=9cm]{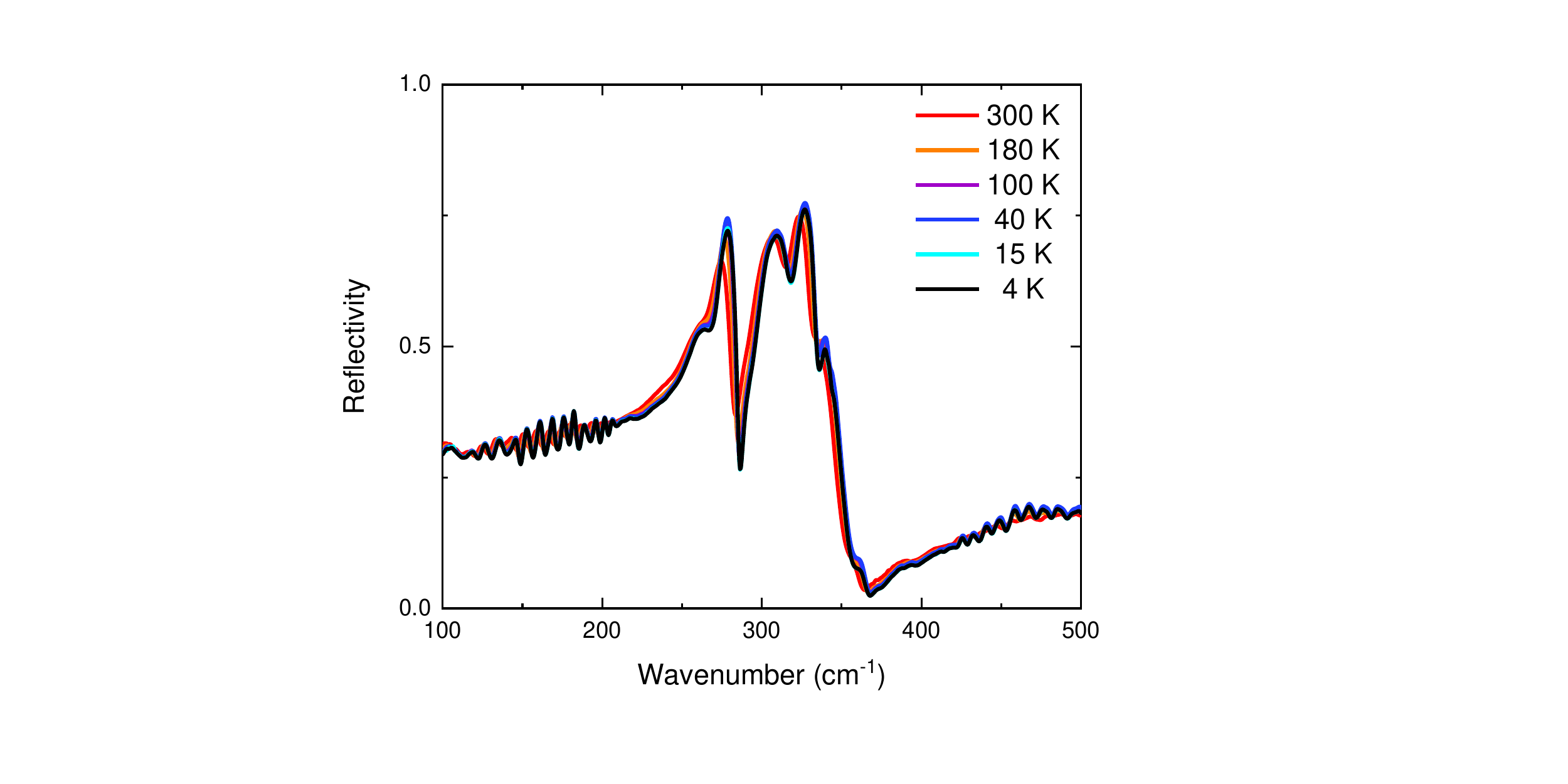}\\
\caption{ Temperature dependence of reflectivity spectra of \ngs\ with $E \parallel (ab)$}\label{fig:IR}
\end{figure}


\begin{table*}
	\caption{Measured frequencies $\omega$ and widths $\gamma$ for the Raman and IR active modes and the polarizations in which they appear. The {\it italic} font is used for those bands, which appear in violation of the selection rules. The bands marked with * appears in \cite{Lutz1986}. Experimental
 frequencies are compared with those determined from calculations, and the relative displacements of each of the unique atomic positions is shown.}
	\label{tab:phocomp}
	\begin{tabular*}{\textwidth}{@{\extracolsep{\fill}}*{11}{c}}
		\\ \hline \hline
			& \multicolumn{3}{c}{Observed Raman bands} & \multicolumn{2}{c}{Observed IR} & \multicolumn{5}{c}{DFT calculations} 	\\	\cline{2-4} \cline{5-6}	\cline{7-11}
		Sym	(pol)	& $\omega$	& $\gamma$	& Observed Pol.		& $\omega$	& Pol.	& $\omega$	& Ni	& Ga	& S1	& S2	\\ \hline
		$A_{2u} (z)$	&		&		&		&		&	& 9		& 0.38	& 0.38	& 0.38	& 0.38	\\
		$E_u (x)(y)$		&		&		&		&		&	& 18		& 0.38	& 0.38	& 0.38	& 0.38	\\
		$E_g (xy)$		& 42		& 5		& $xx,xy,RR$	&		&	& 40		& 0.00	& 0.50 	& 0.49	& 0.10	\\
	    $E_u (x)(y)$		&		&		& 		&		&	& 62		& 0.54 	& 0.30	& 0.28 	& 0.44 	\\
		$A_{1g} (xx)$	& 120		& 5		& $xx,RL$	&		&	& 118		& 0.00	& 0.44	& 0.50	& 0.24	\\
		$A_{2u} (z)$	&		&		&		&		&	& 123		& 0.86	& 0.19	& 0.30	& 0.07	\\
		$E_g (xy)$		& 206		& 19		& $xy,RR$	& {\it 187*}	&	& 198		& 0.00	& 0.06	& 0.00	& 0.70	\\
        	none  		&		& {\it 233} 	& 	$xx$       &	        &       &		&       &       &       &       \\
		$E_u (x)(y)$	& 276		&		&	xx,xy	& 271		& $x,y$	& 261		& 0.54	& 0.11	& 0.16	& 0.56	\\
		$A_{2u} (z)$	& {\it 276}	&		& xx,xy		&		&	& 280 		& 0.12	& 0.08	& 0.52	& 0.46	\\
		$E_g (xy)$		& wing $\sim$ 290		&	xy	& 		&		&	& 284		& 0.00	& 0.28	& 0.65	& 0.05	\\
		$E_u$	& 312		&		&	xx,xy	& 298		& $x,y$	& 284		& 0.14	& 0.27	& 0.62	& 0.17	\\
		$A_{1g}$	& 313		& 10		& $zz,RL$	& {\it 316}		&	& 316		& 0.00	& 0.21	& 0.57	& 0.34	\\
		 none  		&		&	        &	        &           	&  {\it 335} &   	&       &       &       &      \\
	        $A_{2u}$	&{\it 357}		&		& $xx,RL$   & 356{\bf w} 	& unpol.	& 354	& 0.03	& 0.37	& 0.33	& 0.50	\\
		$A_{1g}$	& 450		& 14		& $xx,RL$	&  		&	& 408		& 0.00	& 0.25	& 0.18	& 0.63	\\
                \hline \hline
	\end{tabular*}
\end{table*}

\subsection{Magnetic susceptibility}

\begin{figure}
  \centering
  \includegraphics[width=9cm]{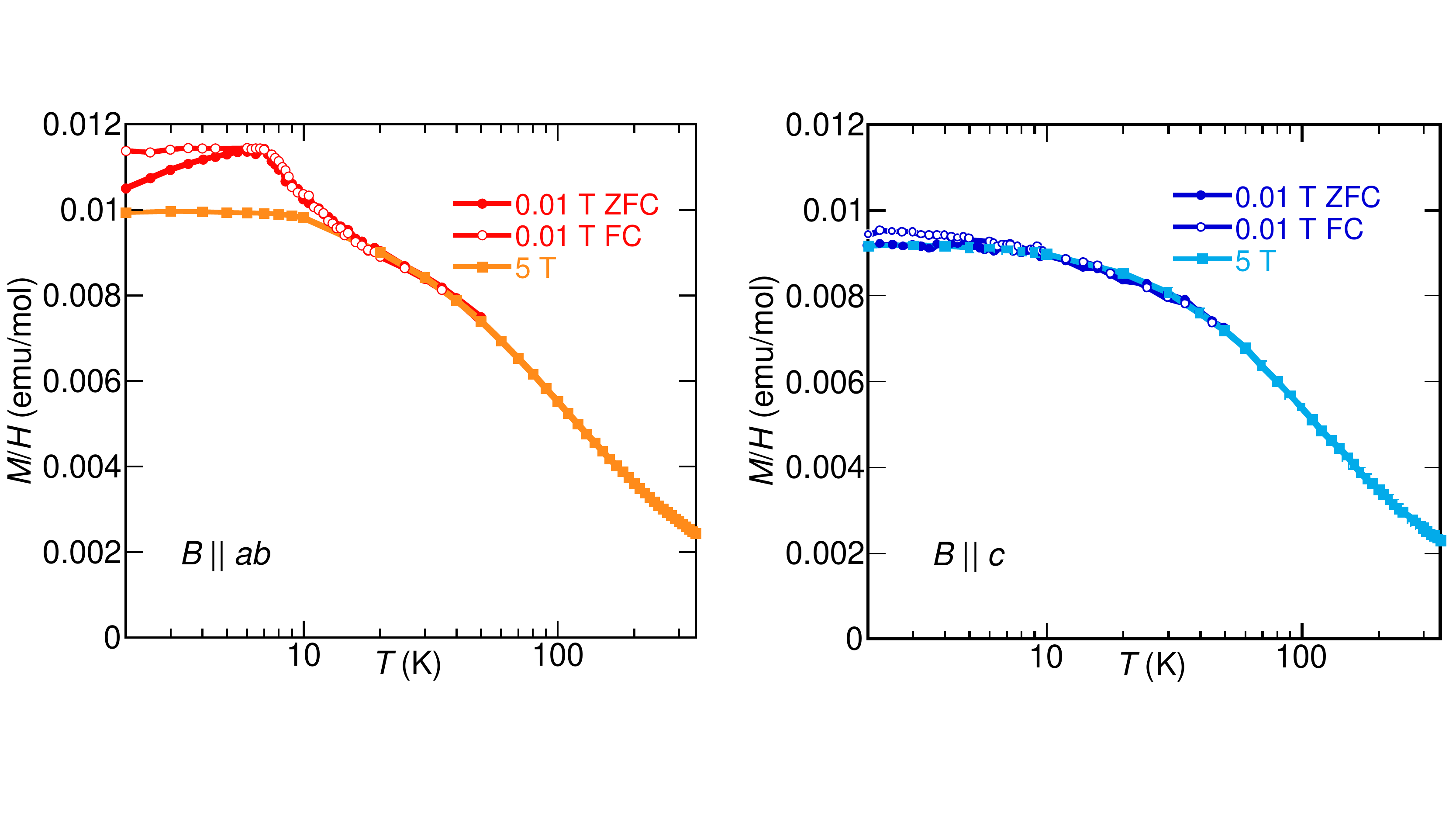}\\
\caption{Temperature dependence of magnetic susceptibility of \ngs\ for the fields applied in the $(ab)$ plane and along the $c$-axis (right panel) at 0.01 and 5 T.}\label{fig:mag}
\end{figure}

The temperature dependence of the susceptibility $\chi(T) = M(T)/H$ at H = 0.01 and 5 T for the in plane ($\chi_{ab}$)) and out-of-plane ($\chi_{c}$) direction is presented in Fig.~\ref{fig:mag} (a) and (b), respectively. Hysteresis between the field-cooled (FC) and zero-field-cooled (ZFC) data under 0.01 T is seen below a freezing temperature $T_{SF}$ = 7 K. The Weiss temperature $\Theta_W$ = −83(1) K and the effective moments $p_{eff}$ = 2.81 $\mu_B$ /Ni are estimated by the Curie-Weiss law, $\Theta_W = C/(T − \Theta_W)$ for temperatures 150 K $\leq$ T $\leq$ 350 K under 5T.

\end{document}